\documentclass[aps,prl,twocolumn]{revtex4}
\usepackage{amsfonts}
\usepackage{amssymb}
\usepackage{epsfig}
\usepackage{float}
\usepackage{amsmath}

\begin{document}

\title{Electronic structure and magnetism in the frustrated antiferromagnet
LiCrO$_{2}$}
\author{I.~I. Mazin}
\affiliation{Code 6391, Naval Research Laboratory, Washington, D.C. 20375}

\begin{abstract}
 LiCrO$_{2}$ is a 2D triangular antiferromagnet, isostructural with the
common battery material LiCoO$_{2}$ and a well-known Jahn-Teller
antiferromagnet NaNiO$_{2}.$ As opposed to the latter, LiCrO$_{2}$ exibits
antiferromagnetic exchange in Cr planes, which has been ascribed to direct
Cr-Cr $d-d$ overlap. Using LDA and LDA+U first principles calculations I
confirm this conjecture and show that (a) direct $d-d$ overlap is indeed
enhanced compared to isostructural Ni and Cr compounds, (b) $p-d$ charge
transfer gap is also enhanced, thus suppressing the ferromagnetic
superexchange, (c) the calculated magnetic Hamiltonian maps well onto the
nearest neighbors Heisenberg exchange model and (d) interplanar inteaction
is antiferromagnetic.
\end{abstract}

\maketitle

The series of compounds with a common formula $ {AM}$O$_{2},$
 where $%
 {A}$ is an alkaline or noble metal, usually Li or Na, and $%
 {M}$ is a $3d$ metal, formed by triangular $ {M}$O$_{2}$
layers stacked hexagonally (\textit{e.g.,} LiCoO$_{2}$) or rhombohedrally (%
\textit{e.g.,} LiNiO$_{2}$) with full or partial intercalation by $ {%
A,}$ has been attracting considerable recent interest, largely driven by the
immense importance of the LiCoO$_{2}$ compound in electrochemical industry
and by the unconventional superconductivity discovered in the hydrated Na$%
_{1/3}$CoO$_{2}.$ The nickelates, LiNiO$_{2},$ NaNiO$_{2},$ AgNiO$_{2},$ Ag$%
_{2}$NiO$_{2}$ have been also subject of numerous studies, mostly because of
their magnetic properties coupled with interesting structural
transformations. However, chromates, such as LiCrO$_{2},$ NaCrO$_{2},$ and
KCrO$_{2},$ despite their potential use for rechargeable batteries\cite{bat1}
and as catalists\cite{catal} have been studied experimentally only
sporadically\cite{French,neutrons,jap,Angel,XPES,Cava}, and no first
principle calculations, to the best of my knowledge, have been reported so
far.

In this paper I report all-electron full-potential electronic structure
calculations for LiCrO$_{2}.$ In agreement with the experimental findings,
the magnetic interaction in-plane is found to be strongly antiferromagnetic.
Interplane magnetic interaction is very weak and also antiferromagnetic. The
total energy calculations 
for three different collinear magnetic
configurations map perfectly well onto the standard nearest-neighbor
Heisenberg model. As conjectured in the first experimental papers\cite{French}
the main reason for switching the in-plane magnetic interactions from
ferromagnetic in LiNiO$_{2}$ to antiferromagnetic in LiCrO$_{2}$ is mainly
 the enhanced direct overlap between the metal $d-$orbitals in chromates,
while, additionally,
the increased charge-transfer $p-d$ gap 
reduces the ferromagnetic superexchange in chromates as well\cite{Danya}. 
Finally I will discuss the
role of Coulomb correlations as revealed by LDA+U calculations.

LiCrO$_{2}$ crystallizes in a rhombohedral R$\overline{3}$m structure\cite%
{neutrons} with the lattice parameters $a=2.898$ \AA , $c=14.423$ \AA , and
with the O height $z_{O}=0.261.$ Its magnetic structure is close to the
ideal 120$^{o}$ structure characteristic of the nearest neighbor Heisenberg
model on a triangular plane. The computational results reported below were
obtained using the standard full-potential linearized augmented plane wave
code WIEN2k\cite{Wien}. Exchange and correlation were taken into acount in a
gradient approximation of Perdew $et$ $al$\cite{PBE}. Convergence with
respect to both the cutoff parameter $RK_{\max }$ and the number of the
inequivalent k-points (up to 400+) was checked.

\begin{figure}[htbp]
\centering
\includegraphics[angle=0,width=0.95\linewidth]{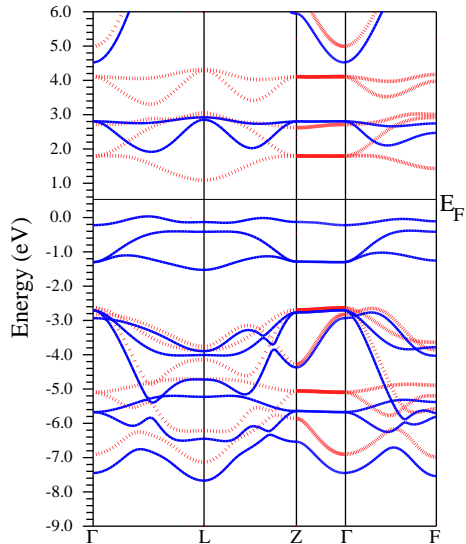}
\caption{ Band structure of LiCrO$_2$ in the hypothetical
ferromagnetic state. Spin-majority bands are solid and blue,
spin-minority ones dotted and red. (color online).  }
\label{FM1bands} \end{figure}

I start first with the hypothetical ferromagnetic structure. Cr $d$-states
are split by the crystal field into 3 $t_{2g}$ and 2 $e_{g}$ states,
separated by roughly 3 eV in the spin-majority and 2 eV in the spin-minority
channel (Fig. \ref{FM1bands}). The difference is due to the fact that the
crystal field due to O-Cr hybridization 
is stronger in the
spin-majority channel where the energy separation between the oxygen $p$%
-states and the Cr $d-$states is smaller. Cr$^{3+}$ has exactly three
electrons, and due to the strong exchange splitting it turnes out to be well insulating already 
in the LDA; thus it can be treated as band insulator.
(I will show later
that the effect of the Mott-Hubbard correlations is relatively small), with
the band gap between the occupied spin-up and unoccupied spin-down $t_{2g}$
bands. The minimum spin-flip gap is slightly larger than 1 eV. The average
spin-flip energy between the two bands is close to 2.9 eV. The oxygen states
are well separated from the $d-$states, as opposed to the isostructural
oxides of the late 3d metals (Co, Ni), where the corresponding bands
essentially overlap. The calculated magnetic moment is, obviously 3 $\mu
_{B},$ not too far from the experimentally measured noncollinear magnetic
moment of 2.68$\pm 0.13$ $\mu _{B}$\cite{jap}. Structural optimization with
experimentally constrainted cell dimensions $a$ and $c$ leads to the O
position $z_{O}=0.258,$ corresponding to an approximately 4\% larger distance between Cr and O planes
compared to the experiment\cite{neutrons}, and a Cr-O-Cr angle of 93$^\circ$
compared to the experimental 94.6$^\circ$. I will discuss later possible
relation of this deviation to magnetic properties. 

LiCrO$_{2}$, being a good isulator already in LDA, should exhibit an antiferromagnetic
superexchange interaction between the CrO$_{2}$ layers, proportional to $%
t_{\perp }^{2}/\Delta _{s.f.},$ where $t_{\perp }$ is the effective
interlayer hopping, a small number since it proceeds $via$ a long path of
Cr-O-O-Cr\cite{OLiO}, and $\Delta _{s.f.}$ is the energy cost for flipping a d 
electron spin (in LSDA, it is Stoner $I$, in the Hubbard model $U$).
 To get an estimate (from above\cite{note1}!) of this
interaction I compare the total energy of the ferromagnetic state and the
A-type antiferromagnetic state (ferromagnetism in plane, antiferromagnetic
stacking). The energy difference appears to be 3$\pm 1$ meV per two formula
units, a very small number indeed. This strongly suggests that LiCrO$_{2}$
is a very good model material for studying frustrated magnetism in two
dimensions. NaCrO$_{2}$ maybe even a better example of the same physics, as
discussed in Ref. \cite{Cava} from an experimental point of view.

Let me now discuss magnetic interaction in plane. 
The above-described exchange splitting that renders LiCrO$_{2}$ a band insulator
in LDA calculations, being essentially a local effect, should be operative in an antiferro-,
as well as in any ferrimagnetic arrangement. In other words, no new magnetic physics is
introduced by Mott-Hubbard effects. Indeed, in the antiferro- and ferrimagnetic 
calculations, described in more details below, I invariably found insulating ground states 
(a corollary of this finding is that the magnetically disordered state above $T_N$
will be also insulating even in LDA).

  The isostructural
compounds with higher transition metals (Co, Ni),
 when order magnetically,
assume the so-called A-type antiferromagnetism, that is, ferromagnetic
planes, stacked antiferromagnetically. The mechanisms for the in-plane
ferromagnetism are well understood: this is the classical 90$
^\circ
$ superexchange, plus, in metallic compounds, like Na$_{0.7}$CoO$_{2},$
Stoner ferromagnetism. LiCrO$_{2}$ in not metallic, however, the O-Cr-O bond
angle is fairly close to 90$^\circ$
 and one expects the corresponding superexchange to be ferromagnetic.

A popular explanation\cite{French} is that this superexchange is indeed
ferromagnetic but is surpassed by direct-overlap exchange between Cr $d$%
-orbitals. It was also pointed out\cite{Danya} that the O $p$ band in the
higher metal (Co, Ni) is located higher with respect to the metal d-band,
which enhances the $p-d$ hybridization and the superexchange, compared to
that in LiCrO$_{2}.$ While logical $per$ $se$, however, neither of these
propositions has been subjected to to a quantitative test. Besides, although among the
physicist dealing with transition metal oxides it is generally known that
direct $d-d$ exchange is antiferromagnetic, this is not a trivial or obvious
statement. Suffice it to remind the reader that essentially all textbooks in
solid state physics describe direct exchange in terms of the original
Heisenberg theory that in fact leads to the \textit{ferromagnetic}
interactions. It is instructive to revisit the issue of the net sign of a
direct exchange interaction from the local spin-density functional (LSDA)
point of view, which is the foundation of all quantitative investigations in
this direction, and I refer the reader to the Appendix where such an analysis 
is presented.

\begin{figure}[htbp]
\centering
\includegraphics[angle=0,width=0.47\linewidth]{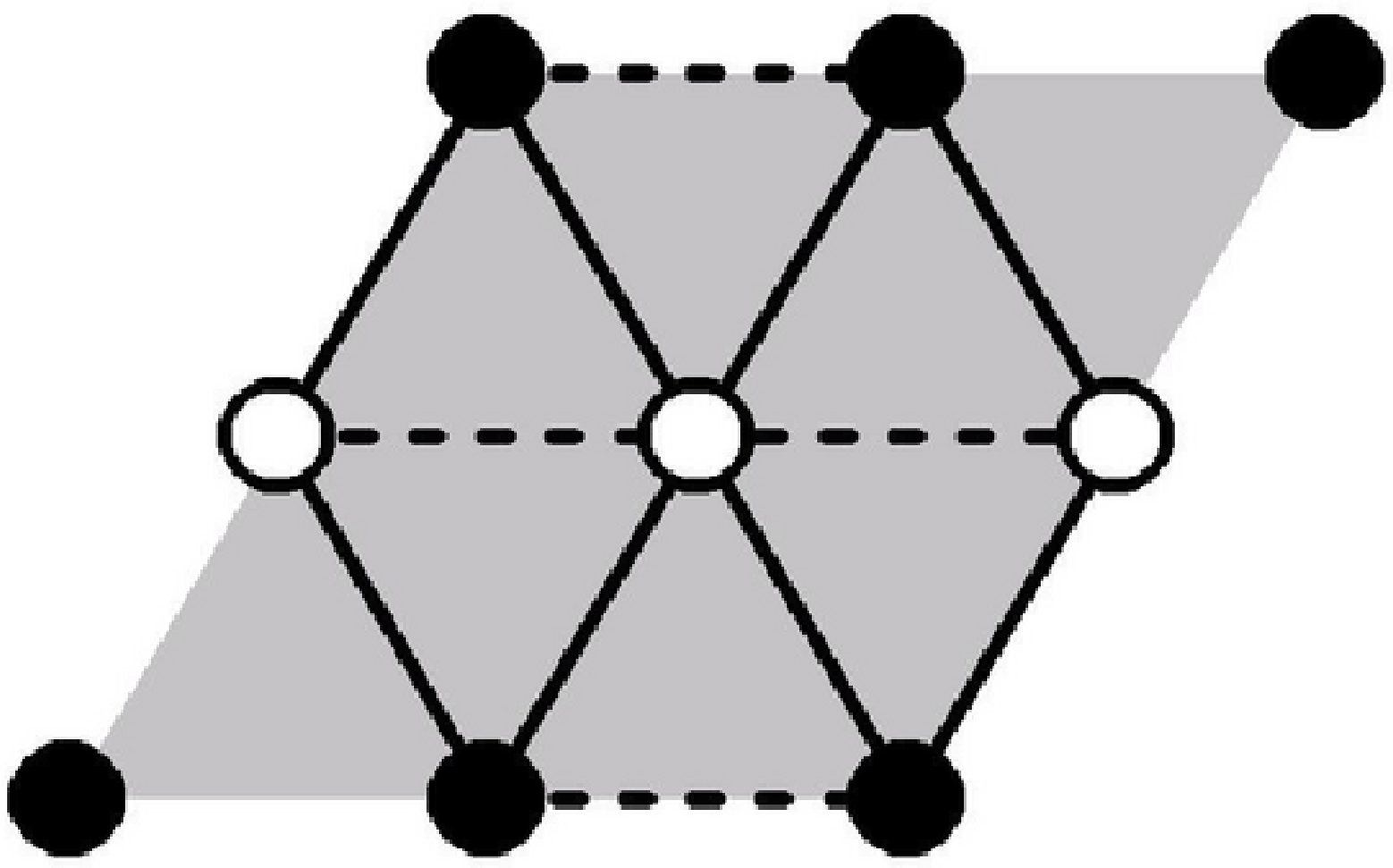}
\includegraphics[angle=0,width=0.47\linewidth]{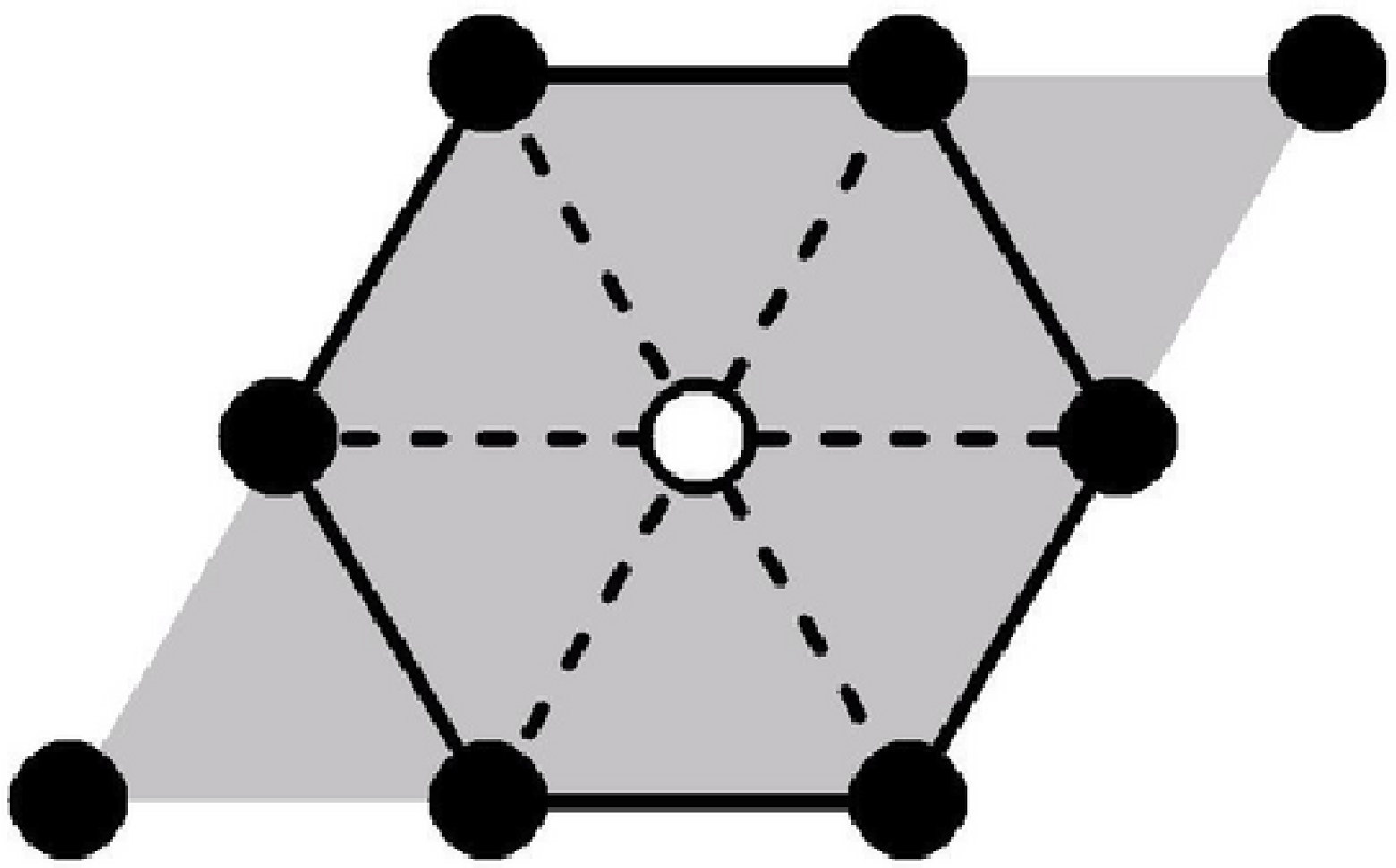}
\caption{Two different magnetic pattern (``antiferro'': left,
``ferrimag: right'') used in calculating the 
exchange constants. Filled (open) circles indicate up (down) moments within
in the supercell, solid (dashed) line ferro- (antiferro-)magnetic bonds. }
\label{str} \end{figure}

To address the nature of the magnetic interactions in LiCrO$_{2}$,
let us compare it with the isostructural LiNiO$_{2}.$ LSDA
calculations predict in the latter a ferromagnetic exchange in plane, of the
order of 5 meV\cite{us}. To compare with LiCrO$_{2}$, I have performed LDA 
calculations in a 2$\times 2 \times 1$ supercell, assuming three different 
magnetic patterns inside the plane: ferromagnetic, antiferromagnetic (alternating 
ferromagnetic chains running along 100, see Fig. \ref{str}), and ferrimagnetic,
relatining the hexagonal symmetry by flipping one spin out of four (Fig. \ref{str}).
The results can be perfect well mapped onto the nearest neighbor Ising model,
yielding  \textit{%
antiferromagnetic }exchange $\gtrsim 20$ meV (22 or 23 meV, depending 
on which two lines in Table 1 are used). This is, of course, in excellent agreement with the
experiment, but how does it answer to the theoretical conjectures described
above?

\begin{table}[tbp]
\caption{LSDA energies with respect to the energy of the FM 
state (in meV), magnetic moments insides the Cr MT spheres
(in $\mu_B$), and excitation gaps (in eV) for
 several magnetic states in a quadruple 2$\times$2
LiCrO$_2$ supercell.}
\begin{tabular}{c|c|c|c|c|c|c}
ordering & Num. of                  & Num. of   & $\bar{M}$ (Cr) & $M_{tot}$ & $E$&gap \\ 
         &Cr$\uparrow $/Cr$\downarrow$& FM/AFM bonds&                &           &\\
\hline
ferromag & 4/0 & 12 & 2.6 & 12 & 0&1.1 \\ 
antiferro & 2/2 & 4/8 & 2.5 & 0 & -360&1.4 \\ 
ferrimag & 3/1 & 6/6 & 2.5 & 6 & -265&1.25
\end{tabular}
\end{table}

To this end, it is instructive to switch to a less accurate, but more
flexible method, Linear MT-orbitals (LMTO), which provides a capability of
eliminating particular radial functions from the basis, and thus assessing
effects of particular orbital overlaps on the bands. I have calculated the
band structure of LiCrO$_{2}$ and LiNiO$_{2}$ removing all oxygen orbitals
from the basis, so that the resulting d-band width is essentially coming
from direct $d-d$ overlaps. The resulting band widths appear to be for LiCrO$%
_{2}$ and LiNiO$_{2},$ respectively, 1.4 eV and 0.5 eV, indicating that
the direct $d-d$ overlap integrals for Cr are three times large than for Ni,
in accord to earlier conjectures\cite{French}. Comparing these numbers with 
the bandwidths in full calculations shows that in LiCrO$%
_{2}$ most of the total bandwidth is due to the direct overlap, while in
 LiNiO$_{2}$ it mainly comes from the indirect hopping $via$ oxygen.
 Note that  in some crude
approximation this means that the direct exchange due to d-d overlap
is an order of magnitude
stronger in LiCrO$_{2}$ compared to LiNiO$_{2}.$

Let us now look at the O-$p$ -- metal $d$ energy separation. Again, in LMTO
there is a gauge that can be readily used: separation between the
corresponding band-center quasiatomic parameters. In LiCrO$_{2}$ this
separation appears to be 5.5 eV, while in LiNiO$_{2}$ it was 1.6 eV, 3.5
times smaller. This leads to a suppression (albeit not by the same factor)
of the ferromagnetic \textit{super}exchange in the former compound, as
proposed by Khomskii\cite{Danya}.
 It is somewhat hard to estimate the actual
reduction of this interaction, because the more diffuse character of the Cr $%
d$ orbitals leads to some enhancement of the $p-d$ hopping compared to the
Ni compounds. Since the $p-d$ overlap is already strong in the nickelate,
this enhancement is not nearly as dramatic as that of the direct $d-d$
hopping, but it should be present. 

The summary of this part is that while in the late 3d metal layered oxides,
such as (Li,Na)NiO$_{2},$ (Li,Na)CoO$_{2},$ the relatively strong
ferromagnetic interactions (90$^{o}$ superexchange, but also Stoner ferromagnetism,
discussed below) is
largely, but not entirely compensated by the AFM superexchange (due to
deviation from 90$^{o})$ and by the direct $d-d$ AFM exchange. 
In the early
3d metal oxides, (Li,Na)CrO$_{2},$ the latter is greatly (up to an order of
magnitide) enhanced, while the former is suppressed.

I will now address the question of the relative importance of several
factors favoring AFM in-plane interactions in the early 3d oxides. Two have
already been mentioned, stronger deviation of the metal-oxugen-metal angle
from 90$^{o}$ ($e.g.,$ in LiCrO$_{2}$ this angle is 94.6$^{o},$ while in
NaNiO$_{2}$ is close to 92$^{o}),$ and stronger direct d-orbitlas overlap.
There is a third important factor: the Stoner magnetism. Indeed, in such a
system as NaNiO$_{2},$ where Ni is a Jahn-Teller ions with an orbital
degeneracy, the Fermi level
falls inside the $e_{g}$ band, which is wider in
the FM case, and, correspondingly, has lower kinetic energy. In other words,
itinerant $e_{g}$ electrons have more freedom to move in the crystals on a
background of parallel magnetic moments (one can also call this a d-d double
exchange; the difference is purely terminological). Note that it does not
matter if a small Jahn-Teller gap opens up in the $e_{g}$ band; as long as
this gap is smaller than or comparable to the band width, kinetic energy
will still favor the FM arrangement. LiCrO$_{2}$ does not display any
orbital degeneracy, thus lacking this contribution to the magnetic interactions.

\begin{figure}[htbp]
\centering
\includegraphics[angle=0,width=0.95\linewidth]{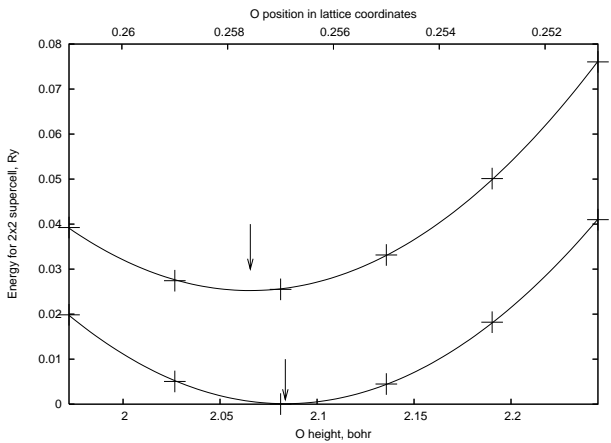}
\caption{Total energy as a function of oxygen position.
The upper curve is for the ferromagnetic, the lower for
the ferrimagnetic case (see Table 1).
The range of positions spans the experimental structure
from Ref.\cite{neutrons} on the left to the structure with
the Cr-O-Cr angle equal to 90$^\circ$ on the right.
Calculated equilibrium positions are indicated by arrows. 
  }
\label{phon} \end{figure}

It is tempting to try to get an idea of the relative importance of the two FM
interactions. I attemted to  address this issue by doing
 calculations for a hypothetical LiCrO$_{2}$ 
 with
oxygen octahedra unsqueezed so as to have Cr-O-Cr angle exactly 90$^{o},
$ while keeping the intraplane Cr-Cr distance constant (otherwise the calculations 
proceeded exactly as for the experimental structure, as described above in Table 1 
and Fig. \ref{str}). One may think that this procedure would substantially enhance
the 90$^{o}$ FM superexchange, hopefully keeping other magnetic interactions unchanged.
 The result seems, on the first glance, unusual: the exchange interaction 
becomes {\it more} antiferromagnetic (by about 17 meV).  A closer look, however,
reveals that this can be traced down to the fact that the equilibrium 
O-Cr bond length in the AFM structure appears to be longer than in the FM structure 
(see Fig.\ref{phon}), providing less hybridization and slightly smaller exchange 
splitting; note that the AFM energy gain due to the Cr-Cr direct exchange 
inversly depends on the exchange splitting.  The total energy curves  
in Fig.  \ref{phon} are practically rigidly shifted with respect to each other:
the curvature is the same within the computational accuracy, corresponding to
the frequency of the A$_{1g}$ phonon of 567 cm$^{-1}$.

\begin{figure}[htbp]
\centering
\includegraphics[angle=0,width=0.95\linewidth]{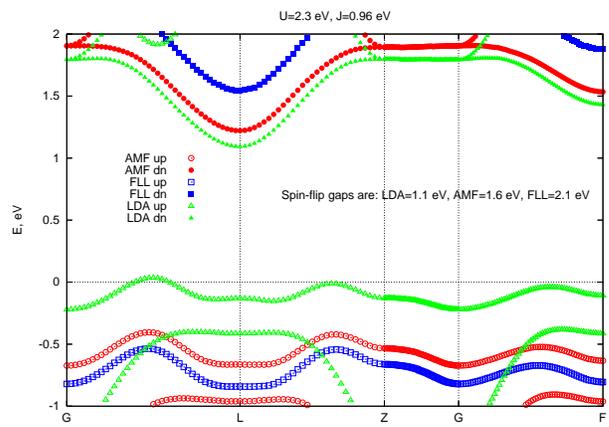}
\caption{ Band structure of ferromagnetic LiCrO$_2$ in LDA
and in two different LDA+U flavors.
 (color online).  }
\label{LDAU} \end{figure}

Finally, although one can expect that the role of the on-site Mott-Hubbard
correlations should be small, given large widths of the Cr d bands, it is of
interest to estimate it using a standard implementation of the
LDA+U technique\cite{Wien}. 
The
latter is most often used in one of the two flavors\cite{Petukhov}: one is
designed to reporoduce the 
fully localized limit (FLL), the other simulates
fluctuations around the mean-field solution (AMF). The former is believed to
be more appropriate for large $U$ system, and the latter for small $U$'s
(admittedly, in this regime the very concept of LDA+U becomes questionable).
I have estimated the parameters using the LMTO internal quasiatomic loop\cite%
{Molodtsov} to be $U=2.3$ eV, $J=0.96$ eV. This is a moderate $U,$ probably
more on the AMF side. Because of that, I have performed LDA+U calculations
in both limits, in order to compare them with each other and with LDA.
 The
results (assuming the FM structure) are shown in Fig. \ref{LDAU}. As expected, LDA+U does not introduce
any new physics: the minimal gap is still the spin-flip gap between $t_{2g}$
and $e_{g},$ 
enhanced by 0.$5$ eV in AMF, due to a shift of the occupied
band by $\approx (U-J)/2$, and by 1 eV in FLL, due to an additional shift of
the unoccupied states by the same amount. Note that these shifts are smaller
than those expected in a typical band insulator due to the well-known
density-derivative discontinuity\cite{ddc}, related to an unscreened long-range exchange
interaction\cite{gap}. For example, for ZnS, which has a gap of 3.8 eV, LDA
gives a gap of 1.7-1.8 eV. While I was not able to locate data for the experimental
gap in LiCrO$_{2}$, the optical gap in NaCrO$_2$ was reported to be about 3.5 eV\cite{opt},
thus making comparison with ZnS rather meaningful. Note that the numbers in Fig. 
\ref{LDAU} are for the FM ordering; introduction of antiferromagnetism of course
increases the gap (cf. Table 1).

To conclude, one can describe LiCrO$_{2}$ as a weakly correlated band
insulator with the minimal gap beng the spin-flip $t_{2g}-e_{g}$ gap.
Magnetically it is a frustrated nearest-neighbor 2D triangular
antiferromagnet, with a sizeable exchange constant of the order of 20 meV.

I would like to acknowledge many useful discussions with Michelle Johannes
and Daniel Khomskii.

\section{Appendix: Direct exchange in LDA}

Consider two overlaping orbitals $\phi _{1}\equiv \phi (r),$ $\phi
_{2}\equiv \phi (|\mathbf{r}-\mathbf{R|})$ centered at the points separated
by $R.$ The corresponding atomic (on-site) energy we shall call $E,$ and we
assume that there is intraatomic exchange splitting such that the spin-up
state has the energy $E,$ but the spin-down state has the energy $E+I.$ The
intraatomic exchange parameter, $I,$ in LSDA is known as \textquotedblleft
Stoner parameter\textquotedblright . For strongly correlated systems Hubbard 
$U>>I$ plays the leading role instead. 
One may expect that in an intermediate regime the
cost of flipping a spin is intermediate between the LSDA $I$ and Hubbard $U.$
The so-called Andersen's force theorem\cite{force} states, among other
things, that in LSDA the total energy difference between the ferro- and
antiferromagnetic arrangement can be computed as the difference in
one-electron energy plus the difference in magnetic (\textquotedblleft
Stoner\textquotedblright ) energy. The one-electron part favors
antiferromagnetism. Indeed, the occupied levels in the ferromagnetic
case are not shifted with respect to the on-sites energy for the spin-up
state, but in the antiferromagnetic case, in the second order perturbation
theory, they shift down by $t^{2}/I$ each, where $t$ is the hopping
integral,
 $t=\left\langle \phi _{1}|-\nabla ^{2}+V_{eff}(\mathbf{r})|\phi
_{2}\right\rangle $ (in Ry units).

 The magnetic part favors ferromagnetism,
since the corresponding energy in LSDA is $E_{St}=-\int Im^{2}(\mathbf{r}%
)/4, $ where $m^{2}(\mathbf{r})$ is the total spin density. For the
ferromagnetic alignment $m(\mathbf{r})=$ $\phi _{1}^{2}+$ $\phi _{2}^{2},$ $%
E_{St}=-\int I(\phi _{1}^{2}+\phi _{2}^{2})^{2}/4.$ For the
antiferromagnetic one $m(\mathbf{r})=$ $\phi _{1}^{2}-$ $\phi _{2}^{2},$ $%
E_{St}=-\int I(\phi _{1}^{2}-\phi _{2}^{2})^{2}/4.$ Thus this term favors
ferromagnetism by $I\left\langle \phi _{1}^{2}|\phi _{2}^{2}\right\rangle $
and is the LSDA countrepart of the textbook direct ferromagnetic exchange.
Note that in transition metals $I$ is on the order of 0.5--1 eV. Assuming
that the tails of the $d-$wave function decay as $\exp (-r/r_{d}),$ we see
that the ferromagnetic exchange is of the order of $I\exp (-R/r_{d}).$ In $%
t, $ for weak overlaps, the main role is played by the kinetic energy, $%
t\sim \left\langle \phi _{1}|-\nabla ^{2}|\phi _{2}\right\rangle \sim
\left\langle \phi _{1}|\phi _{2}\right\rangle /r_{d}^{2}\sim \exp
(-R/2r_{d})/r_{d}^{2}.$ The ratio $J_{FM}/J_{AFM}$ is thus $I^{2}\exp
(-R/r_{d})/[\exp (-R/2r_{d})/r_{d}^{2}]^{2}=$ $I^{2}r_{d}^{4}.$ Recalling
that in solids the internuclear distance $R$ is of the order of the lattice
parameter $a$, while the Fermi vector is of the order of the Brillouin zone
radius, $\pi /a, $ we can estimate $J_{FM}/J_{AFM}\sim
(I^{2}/m_{d}^{2}E_{F}^{2})(\pi r_{d}/a)^{4},$ where $m_{d}\sim 5-10$ is the
effective $d-$band mass, $E_{F}\sim 3-5$ eV is the $d-$band width, $r_{d}$ $%
\sim $ 1 (in Bohr radii), $a\sim 5-10.$ Thus $J_{FM}/J_{AFM}\sim
10^{-3}.$


\begin{thebibliography}{99}
\bibitem{bat1} K.-S. Kim, S.-W. Lee, H.-S. Moon, H.-J. Kim, B.-W. Cho, W.-I.
Cho, J.-B. Choi, J.-W. Park, J. Power Sources, \textbf{129,} 319 (2004); L.
Zhang and H. Noguchi, J. Electrochem. Soc., \textbf{150}, A601 (2003); P.
Arora, D. Zhang, B.N. Popov, R.E. White, Electrochem. Sol. State Lett. 
\textbf{1}, 249 (1998).

\bibitem{catal} E. Cauda, D. Mescia, D. Fino, G. Saracco, and V. Specchia,
Ind. Eng. Chem. Res. \textbf{44}, 9549 (2005).

\bibitem{French} C. Delmas, G. Le Flem, C. Fouassier and P. Hagenmuller, J.
Phys. Chem. Sol. \textbf{29}, 55 (1978).

\bibitem{neutrons} J.L. Soubeyroux, D. Fruchart, C. Delmas, and G. Le Flem,
J. Mag. Mag. Mat. \textbf{14}, 159 (1979).

\bibitem{jap} H. Kadowaki, H. Takei, and K. Motoya, J. Phys. C., \textbf{7},
6869 (1995).

\bibitem{Angel} S. Angelov, J. Darriet, C. Delmas, and G. Le Flem, Sol.
State Comm., \textbf{50}, 345 (1984).

\bibitem{XPES} V.R. Galakhov, E.Z. Kurmaev, St. Uhlenbrock, M. Neumann, D.G.
Kellerman, V.S. Gorshkov, Sol. State Comm., \textbf{95}, 347 (1995)

\bibitem{Cava} A. Olariu, P. Mendels, F. Bert, B. G. Ueland, P. Schiffer, R.
F. Berger, and R. J. Cava, Phys. Rev. Lett. \textbf{97}, 167203 (2006).

\bibitem{Danya} D.I. Khomskii, unpublished.
\bibitem{Wien} P. Blaha \textit{et al.}, WIEN2k, An Augmented Plane Wave +
Local Orbitals Program for Calculating Crystal Properties, Karlheinz
Schwarz, Techn. Universit$\ddot{a}$t Wien, Austria, 2001.

\bibitem{PBE} J. P. Perdew, S. Burke, and M. Enzerhof, Phys. Rev. Lett., 
\textbf{77}, 3865 (1996).

\bibitem{OLiO} The O-O hopping between the layers is partially ehanced by
assisted O-Li-O hopping. Detailed analysis in the isostructural NaCoO$_2$ 
(M.D. Johannes, I. I. Mazin, and D.J. Singh, Phys.Rev. {\bf B71}, 214410,
2005) shows that both direct and indirect paths yield comparable contributions.

\bibitem{note1} For ferromagnetically ordered layers the effective
interlayer hopping is considerably enhanced because hopping to six second
neighbors in the next layer is comparable to the hopping to the first
neighbor located right above the atom in question (see M.D. Johannes, I. I.
Mazin, and D.J. Singh, Phys.Rev. \textbf{B71}, 214410, 2005). In-plane
antiferromagnetism suppresses already weak interplanar hopping.

\bibitem{force}A. K. Mackintosh and O. K. Andersen, in 
{\it Electrons at the Fermi Surface}, ed. M. Springford 
(Cambridge University Press, Cambridge, 1975).
\bibitem{us} M.J. Johannes, I.I. Mazin and N. Bernstein, unpublished.
\bibitem{Petukhov}A.G. Petukhov, I.I.Mazin, L. Chioncel and A. I. Lichtenstein, 
Phys. Rev. {\bf B67}, 153106 (2003).
\bibitem{Molodtsov}I. I. Mazin and S.L. Molodtsov, Phys.Rev. {\bf B72}, 172504
(2005).
\bibitem{ddc}
J. P. Perdew and M. Levy, Phys. Rev. Lett. {\bf 51}, 1884 (1983);
 L. J. Sham and M. Schlüter, {\it ibid.}, p. 1888.
\bibitem{gap}E.G.Maksimov, I.I.Mazin, S.Y.Savrasov, iand Y.A.Uspenski, J. Phys. Cond. Matt, 
{\bf 1}, 2493 (1989).
\bibitem{opt} P.R. Elliston, F. Habbal, N. Saleh, G.E. Watson,
K.W. Blazey, and H. Rohrer, J. Phys. Chem. Sol. {\bf 36}, 877 (1975).
\end{thebibliography}
\end{document}